\documentclass[submission,copyright]{eptcs}
\providecommand{\event}{GRAPHITE 2013} 
\usepackage{breakurl}             

\usepackage{amsmath}
\usepackage{amssymb}
\usepackage{gastex}
\usepackage{float}
\usepackage{subfigure}

\usepackage{todonotes}

\newtheorem{theorem}{Theorem}

\newtheorem{definition}{Definition}
\newtheorem{remark}{Remark}
\newtheorem{example}{Example}

\newtheorem{proof}{Proof}

\newcommand{\sys}{Sys}
\newcommand{\im}{IM}          
\newcommand{\comp}{K}         
\renewcommand{\int}{Int}  

\newcommand{\hl}[1]{\textbf{#1}}

\newcommand{\tr}[1]{{\xrightarrow{#1}}}

\title{Reachability in Cooperating Systems with Architectural Constraints is PSPACE-Complete} 
    
  \author{Mila Majster-Cederbaum \qquad\qquad Nils Semmelrock\thanks{Corresponding author - Phone: (49) 621-181 2564 - Fax: (49) 621-181 2560}
    \institute{University Mannheim\\ Mannheim, Germany}
    \email{mcb@informatik.uni-mannheim.de \qquad\qquad nsemmelr@informatik.uni-mannheim.de}
  }

  \def\titlerunning{Reachability in Cooperating Systems}
  \def\authorrunning{M. Majster-Cederbaum, N. Semmelrock}

\begin{document}

  \maketitle

  \begin{abstract}
    The reachability problem in cooperating systems is known to be PSPACE-complete. We show here that this problem remains PSPACE-complete when we restrict the communication structure between the subsystems in various ways. For this purpose we introduce two basic and incomparable subclasses of cooperating systems that occur often in practice and provide respective reductions. The subclasses we consider consist of cooperating systems the communication structure of which forms a line respectively a star.
  \end{abstract}
    

  \section{Introduction}

    Cooperating systems are systems that consist of subsystems which cooperate by a \textit{glue-code}. The reachable state space of such systems can be exponentially large in the number of subsystems what is referred to as the \textit{state space explosion problem}. Moreover, there are PSPACE-completeness results for the reachability problem in various formalisms that model cooperating systems, e.g., see \cite{everythingIsPSPACE} for interaction systems, \cite{JonesLL77} for results in $1$-conservative Petri nets and \cite{esparza} for results in $1$-save Petri nets. Clearly, all methods that rely on the exploration of the reachable state space of cooperating systems suffer from these results. Particularly formal verification techniques as LTL or CTL model checking have a runtime that is exponential in the number of subsystems. There are various ways to cope with this problem. One approach is to identify subclasses for which an analysis can be achieved in polynomial time. Hence, the question arises whether there are relevant subclasses of cooperating systems where the reachability problem is decidable in polynomial time. Popular decision problems that are complete in NP or even in PSPACE are decidable in polynomial or linear time in certain subclasses of instances. Maybe the most popular example is the Boolean satisfiability problem where $3$SAT is NP-complete, $2$SAT is decidable in polynomial time and HORNSAT (the problem of deciding whether a given set of propositional Horn clauses is satisfiable) is even decidable in linear time. Similarly, the quantified $3$SAT problem is PSPACE-complete, whereas the quantified $2$SAT problem is also decidable in polynomial time (see \cite{complexGuide} for descriptions and more examples).
    
    There are various starting points to specify subclasses of cooperating systems.
    \begin{enumerate}
      \item Restrictions regarding the behavior of the subsystems.
      \item The degree of synchronization among the subsystems as systems with a very high degree of synchronization tend to display a smaller reachable state space. 
      \item The glue-code, i.e., the structure of the interaction among the subsystems. 
    \end{enumerate}
    Here, our concern lies on the latter.

    As a formal model we consider here interaction systems \cite{interactionSystems}, a very general formalism for modeling cooperating systems that allows for multiway interactions between subsystems. The results in this paper can be easily applied to other formalisms that model cooperating systems. This can be achieved by either adapting the results, e.g., the formalism of interface automata \cite{deAlfaro01} comes close to interaction systems, or by using a mapping among formalisms, e.g., a mapping between interaction systems and $1$-save Petri nets can be found in \cite{christoph08}.
    
    Deciding the reachability problem in general interaction systems is PSPACE-complete \cite{everythingIsPSPACE}. Here we strengthen this result by showing that the reachability problem remains PSPACE-complete in subclasses consisting of interaction systems the communication structure of which forms a star or a linear sequence of components. As star structures appear in practice in, e.g., client/server systems as banking or booking systems and linear structures appear in, e.g., pipeline systems as instruction pipelines or general queue based algorithms, it is important to know that even for such ``simply'' structured systems there is no general efficient analysis method. Our results justify investigations that search for sufficient conditions that can check and guarantee reachability in polynomial time. Also approaches that guarantee correctness by construction, i.e., modeling rules that ensure certain system properties, and are based on structural restrictions become justified by our results. See for example \cite{treeLike,HJK08,Lambertz09,Lambertz11,hoare,roscoeArch,bernardoArch} for approaches that treat these topics. 
    
    The paper is organized as follows. Section \ref{is} contains the definitions. In Section \ref{lin} we introduce a reduction from the acceptance problem in linear bounded Turing machines to the reachability problem in linear interaction systems. A reduction from the reachability problem in general interaction systems to star-like interaction systems is introduced in Section \ref{star}. Section \ref{con} concludes this paper.
    
  \section{Interaction Systems}\label{is}
    
    Interaction systems have been proposed by Sifakis and G\"{o}ssler in \cite{interactionSystems} to model cooperating systems. The model was studied, e.g., in \cite{treeLike,sifakis10,goessler11,moe09,GGM+07}. An interaction system consists of components which cooperate through so called interactions. An interaction specifies a multiway cooperation among components by connecting different interfaces (called ports) of different components. The model is defined in two layers. The first layer, the interaction model, specifies the components, their interfaces and the communication between them. The second layer, the interaction system, describes the behavior of the components by labeled transition systems. In contrast to \cite{interactionSystems} we allow an interaction to be contained in another interaction and do not consider complete interactions.
    
    \begin{definition}\label{defIM}
      Let $K$ be a set of \hl{components} and $\{A_i\}_{i\in K}$ a family of pairwise disjunct sets of \hl{ports} of the components in $K$. In the following we assume that $K=\{1,2,\dots,n\}$. An \hl{interaction} $\alpha$ is a nonempty set of ports from different components, i.e., $\alpha\subseteq\bigcup_{i\in K}A_i$ and for all $i\in K$ $|\alpha\cap A_i|\leq 1$ holds.
      
      An interaction $\alpha_i=\{a_{i_1},a_{i_2},\dots,a_{i_k}\}$ with $a_{i_j}\in A_{i_j}$ ($j\in\{1,2,\dots,k\}$) denotes a possible cooperation among the components $i_1,\dots,i_k$ via their respective ports. A set $Int$ of interactions is called \hl{interaction set (for $K$)}, if each port appears in at least one interaction in $Int$, i.e., $\bigcup_{i\in K}A_i=\bigcup_{\alpha\in Int}\alpha$. The tuple $IM=(K,\{A_i\}_{i\in K},Int)$ is called \hl{interaction model} if $Int$ is an interaction set for $K$.
    \end{definition}
    
    \begin{example}
      \label{ex1}
      Let $r>0$ be a natural number and $K=\{S,c_1,c_2,\dots,c_r\}$ a set of components. $S$ models a server with a set of ports $A_S=\{connect,disconnect\}$ where $connect$ models the connection of a client to this server and $disconnect$ models the disconnection. For $1\leq i\leq r$ component $c_i$ models a client with a set of ports $A_{c_i}=\{connect_i,disconnect_i\}$. $connect_i$ models the connection of client $i$ to the server and $disconnect_i$ the disconnection.
      
      For $1\leq i\leq r$ the interaction $connect\_S\_c_i=\{connect,connect_i\}$ models a connection from client $i$ to the server and the interaction $disconnect\_S\_c_i=\{disconnect,disconnect_i\}$ models the disconnection. Let
      \begin{displaymath}
        \int = \{connect\_S\_c_i,disconnect\_S\_c_i|1\leq i\leq r\}
      \end{displaymath}
      be a set of interactions. Note that $\int$ is an interaction set for $K$, i.e., $\im=(K,\{A_i\}_{i\in K},\int)$ is a well defined interaction model.
    \end{example}

    \begin{definition}
      Let $IM=(K,\{A_i\}_{i\in K},Int)$ be an interaction model. $Sys=(IM,\{T_i\}_{i\in K})$ is called \hl{interaction system} where $T_i=(Q_i,A_i,\to_i,q_i^0)$ for $i\in K$ is a labeled transition systems that models the behavior of component $i\in K$. $Q_i$ is a finite state space, $\to_i\subseteq Q_i\times A_i\times Q_i$ a transition relation and $q_i^0\in Q_i$ an initial state. We refer to $T_i$ for $i\in K$ as the \hl{local behavior} of component $i$ and we denote $q_i\tr{a_i}_iq_i'$ instead of $(q_i,a_i,q_i')\in \to_i$.
      
      We say a state $q_i\in Q_i$ \hl{enables} $a_i$ if there is $q_i'\in Q_i$ with $q_i\tr{a_i}_iq_i'$. We denote the set of enabled ports of a state $q_i\in Q_i$ by $en(q_i)$. This is, $en(q_i) = \{a_i\in A_i|\exists_{q_i'\in Q_i}q_i\tr{a_i}_iq_i'\}$.
    \end{definition}
    
    \begin{example}
      \label{ex2}
      Let $\im=(K,\{A_i\}_{i\in K},\int)$ be the interaction model from example \ref{ex1}. Figure \ref{clse} depicts possible local behavior $T_i$ for $i\in K$. This is, the tuple $Sys=(IM,\{T_i\}_{i\in K})$ is a well defined interaction system. We mark initial states by an incoming arrow.
      \begin{figure}[H]
        \setlength{\unitlength}{1.0cm}
        \begin{center}
          \subfigure[$T_{c_i}$, $1\leq i\leq r$]{
            \begin{picture}(2.0,2.0)(-0.5,0.0)
              \gasset{Nw=0.4,Nh=0.4,Nmr=3}
              \node[Nmarks=i,iangle=90,ilength=0.5](A)(0.5,1.5){}
              \node(B)(0.5,0.5){}
              \drawedge[curvedepth=0.25](A,B){\scriptsize $connect_i$}
              \drawedge[curvedepth=0.25](B,A){\scriptsize $disconnect_i$}
            \end{picture}
          }
          \hspace*{2.5cm}
          \subfigure[$T_s$]{
            \begin{picture}(1.0,2.0)
              \gasset{Nw=0.4,Nh=0.4,Nmr=3}
              \node[Nmarks=i,iangle=90,ilength=0.5](A)(0.5,1.5){}
              \node(B)(0.5,0.5){}
              \drawedge[curvedepth=0.25](A,B){\scriptsize $connect$}
              \drawedge[curvedepth=0.25](B,A){\scriptsize $disconnect$}
            \end{picture}
          }
        \end{center}
        \vspace*{-0.5cm}
        \caption{Local behavior of the components in a simple client/server model.}
        \label{clse}
      \end{figure}
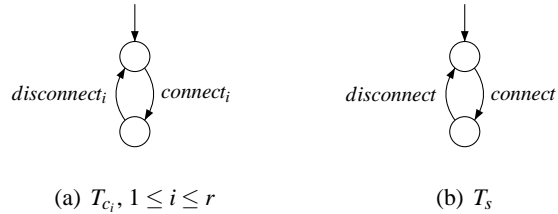
    \end{example}
    
    The behavior of an interaction system is defined as follows.

    \begin{definition}
      Let $Sys=(IM,\{T_i\}_{i\in K})$ be an interaction system where the interaction model is given by $IM=(K,\{A_i\}_{i\in K},Int)$. The \hl{global behavior} of $Sys$ is the transition system $T=(Q,Int,\to,q^0)$ where
      \begin{itemize}
        \item the Cartesian product $Q=\prod_{i\in K}Q_i$ is the \hl{global state space} which we assume to be order independent,
        \item $q^0=(q_1^0,\dots,q_n^0)$ is the \hl{global initial state} and
        \item $\to\subseteq Q\times Int\times Q$ is the \hl{global transition relation} with $q\tr{\alpha}q'$ if for all $i\in K$:
        \begin{itemize}
          \item $q_i\tr{a_i}_iq_i'$ if $\alpha\cap A_i=\{a_i\}$ and 
          \item $q_i=q_i'$ if $\alpha\cap A_i=\emptyset$.
        \end{itemize}
      \end{itemize}
      A state $q\in Q$ is called a \hl{global state}. Globally, a transition $q\tr{\alpha}q'$ can be performed if each port in $\alpha$ can be performed in the state of the local behavior of its respective component.
    \end{definition}
    
    \begin{definition}
      Let $Sys$ be an interaction system and $T=(Q,\int,\to,q^0)$ the associated global transition system. A global state $q\in Q$ is called \hl{reachable} iff there is a path in $T$ that leads from the initial state $q^0$ to $q$. Given an interaction system $\sys$ and a global state $q$, the \hl{reachability problem} consists of deciding whether or not $q$ is reachable in the global behavior of $\sys$.
    \end{definition}
    
    In order to define subclasses of interaction systems we study architectural constraints with respect to the communication structure between components, i.e., our constraints are defined on the interaction model and are independent from the behavior of the components. The communication structure is defined by an undirected graph the nodes of which are components that are connected by an edge if these components are able to interact.
    
    \begin{definition}
      Let $\im=(K,\{A_i\}_{i\in K},\int)$ be an interaction model with $|K|=n$. The \hl{interaction graph} $G=(K,E)$ of $\im$ is an undirected graph with $\{i,j\}\in E$ ($i\not=j$) iff there is an interaction $\alpha\in\int$ with $\alpha\cap A_i\not=\emptyset$ and $\alpha\cap A_j\not=\emptyset$, i.e., if there is an interaction in which both components participate.
      
      An interaction model $\im$ is called \hl{star-like} iff $G$ is a star, i.e., exactly one node is of degree $n-1$ and all other nodes are of degree $1$. $IM$ is called \hl{linear} iff $G$ is connected, two nodes are of degree $1$ and any other node is of degree $2$. An interaction system $\sys$ is called star-like respectively linear if the interaction model of $\sys$ is star-like respectively linear.
    \end{definition}
    
    \begin{remark}
      Note that star-like and linear interaction systems with a set $\int$ of interactions imply that for all $\alpha\in \int$ $|\alpha|\leq 2$.
      
      A star-like or linear interaction system can be seen as a system with a simple hierarchical communication structure, e.g., the simple client/server system in Example \ref{ex1}. Of course, such systems can be far more complex and thus exhibit a highly branched communication structure. This is, a PSPACE-completeness result for deciding the reachability problem in the subclass of star-like or linear interaction systems implies the PSPACE-completeness of deciding the reachability problem in systems with a hierarchical communication structure.
      
      As deciding the reachability problem in general interaction systems is in PSPACE it follows that the same holds for the classes of linear and star-like systems.
    \end{remark}
    
    \begin{example}
      \label{ex3}
      The interaction graph $G$ of the interaction model $\im=(K,\{A_i\}_{i\in K},\int)$ from Example \ref{ex1} is depicted in Figure \ref{igcs}. The interaction graph is a star, i.e., $\im$ is star-like and thus, every interaction system, particularly $\sys$ in Example \ref{ex2}, that contains $\im$ is star-like.
      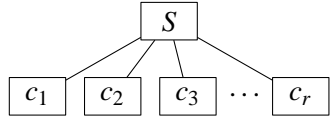
\begin{figure}[H]
        \begin{center}
          \setlength{\unitlength}{1.0cm}
          \begin{picture}(3.5,1.25)
            \gasset{Nw=0.75,Nh=0.5,Nmr=0,AHnb=0}
            \node(A)(1.75,1.0){$S$}
            \node(B)(0.0,0.0){$c_1$}
            \node(C)(1.0,0.0){$c_2$}
            \node(D)(2.0,0.0){$c_3$}
            \put(2.55,0.0){\dots}
            \node(E)(3.5,0.0){$c_r$}
            \drawedge(A,B){}
            \drawedge(A,C){}
            \drawedge(A,D){}
            \drawedge(A,E){}
          \end{picture}
        \end{center}
        \caption{Interaction graph $G$ for the interaction model $\im$ in Example \ref{ex1}.}
        \label{igcs}
      \end{figure}
    \end{example}
    
    \begin{example}\label{exline}
      This example illustrates a linear interaction system. We consider a simple communication pipeline consisting of $n$ stations. Station one initiates passing a message to station two, station two passes the message to station three and so on. If the message arrives at station $n$ then station $n$ passes an acknowledge message, on the same way, back to station one.
      
      Let $\im=(K,\{A_i\}_{i\in K},\int)$ be the interaction model with components $K=\{s_1,s_2,\dots,s_n\}$ for $n\geq 2$ where $s_i$ models station $i$ for $1\leq i\leq n$. A station $s_i$ with $1<i<n$ can receive a message ($rec\_m_i$), pass the message forward ($send\_m_i$), receive an acknowledge ($rec\_a_i$) and pass the acknowledge forward ($send\_a_i$). Station $s_1$ can only send the initial message and receive the acknowledge and station $s_n$ can only receive a message and send an acknowledge. This is, the port sets of the components are defined as follows.
      \begin{displaymath}
        \begin{array}{lcl}
          A_{s_1} &=& \{send\_m_1,rec\_a_1\} \\
          A_{s_i} &=& \{rec\_m_i,send\_m_i,rec\_a_i,send\_a_i\}, \; 1<i<n \\
          A_{s_n} &=& \{rec\_m_n,send\_a_n\}
        \end{array}
      \end{displaymath}
      The interaction set $\int$ is given by the following interactions.
      \begin{displaymath}
        \begin{array}{lcl}
          send\_message_i &=& \{send\_m_i,rec\_m_{i+1}\}, \; 1\leq i<n \\
          send\_acknowledge_i &=& \{send\_a_i,rec\_a_{i-1}\}, \; 1<i\leq n
        \end{array}
      \end{displaymath}
      Let $Sys=(IM,\{T_i\}_{i\in K})$ be the interaction system with local behavior depicted in Figure \ref{pipebeh}.
      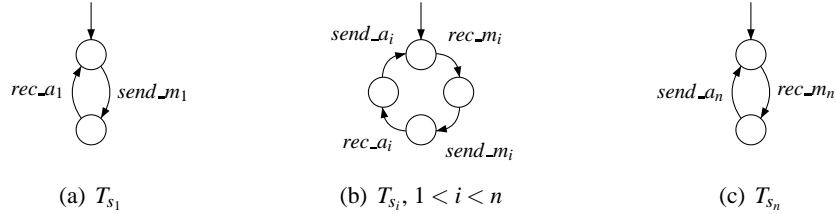
\begin{figure}[H]
        \setlength{\unitlength}{1.0cm}
        \begin{center}
          \subfigure[$T_{s_1}$]{
            \begin{picture}(1.0,2.0)
              \gasset{Nw=0.4,Nh=0.4,Nmr=3}
              \node[Nmarks=i,iangle=90,ilength=0.5](A)(0.5,1.5){}
              \node(B)(0.5,0.5){}
              \drawedge[curvedepth=0.25](A,B){\scriptsize $send\_m_1$}
              \drawedge[curvedepth=0.25](B,A){\scriptsize $rec\_a_1$}
            \end{picture}
          }
          \hspace*{2.5cm}
          \subfigure[$T_{s_i}$, $1<i<n$]{
            \begin{picture}(2.0,2.25)(-0.5,0.0)
              \gasset{Nw=0.4,Nh=0.4,Nmr=3}
              \node[Nmarks=i,iangle=90,ilength=0.5](A)(0.5,1.5){}
              \node(B)(1.0,1.0){}
              \node(C)(0.5,0.5){}
              \node(D)(0.0,1.0){}
              \drawedge[curvedepth=0.25](A,B){\scriptsize $rec\_m_i$}
              \drawedge[curvedepth=0.25](B,C){\scriptsize $send\_m_i$}
              \drawedge[curvedepth=0.25](C,D){\scriptsize $rec\_a_i$}
              \drawedge[curvedepth=0.25](D,A){\scriptsize $send\_a_i$}
            \end{picture}
          }
          \hspace*{2.5cm}
          \subfigure[$T_{s_n}$]{
            \begin{picture}(1.0,2.0)
              \gasset{Nw=0.4,Nh=0.4,Nmr=3}
              \node[Nmarks=i,iangle=90,ilength=0.5](A)(0.5,1.5){}
              \node(B)(0.5,0.5){}
              \drawedge[curvedepth=0.25](A,B){\scriptsize $rec\_m_n$}
              \drawedge[curvedepth=0.25](B,A){\scriptsize $send\_a_n$}
            \end{picture}
          }
        \end{center}
        \vspace*{-0.5cm}
        \caption{Local behavior of the components in a simple communication pipeline.}
        \label{pipebeh}
      \end{figure}
      The interaction graph $G$ of $\im$ is depicted in Figure \ref{igline}. $G$ forms a line of components. Thus, $\im$ is a linear interaction model and $\sys$ is a linear interaction system.
      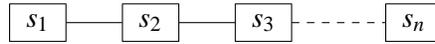
\begin{figure}[H]
        \begin{center}
          \setlength{\unitlength}{1.0cm}
          \begin{picture}(5.0,0.75)
            \gasset{Nw=0.75,Nh=0.5,Nmr=0,AHnb=0}
            \node(A)(0.0,0.0){$s_1$}
            \node(B)(1.5,0.0){$s_2$}
            \node(C)(3.0,0.0){$s_3$}
            \node(D)(5.0,0.0){$s_n$}
            \drawedge(A,B){}
            \drawedge(B,C){}
            \drawedge[dash={0.1}0](C,D){}
          \end{picture}
        \end{center}
        \caption{Interaction graph $G$ for the interaction model $\im$ in Example \ref{exline}.}
        \label{igline}
      \end{figure}
    \end{example}

  \section{PSPACE-completeness of Reachability in Linear Systems}\label{lin}
  
    In the following we give a reduction from the accepting problem in linear bounded Turing machines to the reachability problem in linear interaction systems. We use the following syntax for a Turing machine but we refrain from repeating the well known semantics (see \cite{complexGuide} for details).

    \begin{definition}
      A $4$-tuple $M=(\Gamma,\Sigma,P,\delta)$ is called \hl{deterministic Turing machine} (\hl{DTM}) where
      \begin{itemize}
        \item $\Gamma$ is a finite set of \hl{tape symbols},
        \item $\Sigma\subseteq\Gamma$ is a set of \hl{input symbols} with a distinguished \hl{blank symbol} $b\in\Gamma\setminus\Sigma$,
        \item $P$ is a finite set of \hl{states}, including an \hl{initial state} $p^0$ and two \hl{halt states} $p^Y$ and $p^N$ and
        \item $\delta$ is the \hl{transition function} with $\delta:(P\setminus\{p^Y,p^N\})\times\Gamma\to P\times\Gamma\times\{-1,+1\}$.
      \end{itemize}

      We consider a both-sided infinite tape with cells labeled by integers. Given an input $x\in\Sigma^*$ written on the cells labeled $1$ through $|x|$ we assume $M$ to be initially in the initial state $p^0$ and the tape head pointing at cell $1$. For a string $x\in\Sigma^*$ with $|x|=n$ we denote the $i$th letter in $x$ by $x^i$ for $1\leq i\leq n$.

      A DTM $M$ is called \hl{linear bounded} if no computation on $M$ uses more than $n+1$ tape cells, where $n$ is the length of the input string. A \hl{configuration} of a bounded DTM $M$ is denoted by $(p;\gamma_0,\dots,\underline{\gamma_i},\dots,\gamma_{n+1})$ where $M$ is in state $p$, $\gamma_j$ is the tape symbol in cell $0\leq j\leq n+1$ and the tape head is on cell $i$.
    \end{definition}

    \begin{definition}
      The problem \hl{linear space acceptance} (\hl{LSA}) has as input a linear bounded DTM $M$ and a finite string $x$ over the input alphabet of $M$. The question is whether $M$ accepts $x$, i.e., does $M$ halt in the state $p^Y$. It is well known that LSA is PSPACE-complete \cite{complexGuide}.
    \end{definition}
    
    The idea for our reduction is to model the cells of a DTM $M$ by components of an interaction system $\sys_M$ and the transition function of $M$ by interactions such that a path in the global behavior of $\sys_M$ corresponds to an execution of $M$. In order to calculate the next configuration of $M$ we need the current tape head position, the current tape symbol in the respective cell and the current state of $M$. We model all these informations in each cell, i.e., in order to model the calculation of the next configuration we need interactions between the component that models the cell with the tape head and the respective components that model the neighboring cells.

    Let $M=(\Gamma,\Sigma,P,\delta)$ be a linear bounded DTM and $x\in\Sigma^*$ an input with $|x|=n$. Let $\sys_M=(\im_M,\{T_i\}_{i\in \comp})$ be an interaction system with interaction model $\im=(\comp,\{A_i\}_{i\in \comp},\int)$ such that $\comp=\{0,\dots,n+1\}$.

    The set of ports $A_i$ for a component $i\in K$ with $1\leq i\leq n$ is given by
    \begin{displaymath}
      A_i = \{(p,\gamma)_i^1,(p,\gamma)_i^2|p\in P\setminus\{p^Y,p^N\},\gamma\in\Gamma\}.
    \end{displaymath}
    $(p,\gamma)_i^1$ models that the tape head moves away from cell $i$ where $\gamma$ is the current tape symbol in this cell and $M$ is in state $p$. Analogously, $(p,\gamma)_i^2$ models that the tape head moves onto cell $i$ where $\gamma$ is written and $M$ is in state $p$.
    
    Because of $M$ being linear bounded, we know that $\delta$ does not move the tape head from cell $0$ to the left respectively from cell $n+1$ to the right. Thus, we can omit ports in $A_0$ and $A_{n+1}$ that model a head movement from or onto cell $-1$ and $n+2$. $A_0$ is given by
    \begin{eqnarray*}
      A_0 &=& \{(p,\gamma)_0^1|p\in P\setminus\{p^Y,p^N\},\gamma\in\Gamma,\neg\exists_{p',\gamma'}\delta(p,\gamma)=(p',\gamma',-1)\} \cup \\
        && \{(p,\gamma)_0^2|p\in P\setminus\{p^Y,p^N\},\gamma\in\Gamma,\neg\exists_{p',\gamma'}\delta(p,\gamma)=(p',\gamma',1)\}.
    \end{eqnarray*}
    $A_{n+1}$ is defined analogously. The set of interactions is given by
    \begin{displaymath}
      \int=\{\{(p,\gamma)_i^1,(p,\gamma)_{i+T}^2\}|\exists_{p',\gamma'}\delta(p,\gamma)=(p',\gamma',T),0\leq i+T\leq n+1\}.
    \end{displaymath}

    For $i\in \comp$ let $T_i=(Q_i,A_i,\to_i,q_i^0)$ be the local behavior of component $i$ with $Q_i=\{(p,\gamma)|p\in P\cup\{s\},\gamma\in\Gamma\}$ where $s$ is an auxiliary symbol that is not included in $P$. $(p,\gamma)\in Q_i$ with $p\not=s$ models that the tape head is currently on cell $i$ and the current tape symbol in this cell is $\gamma$. $(s,\gamma)$ models that $\gamma$ is the content of cell $i$ and the tape head is not on this cell. The local initial states are derived from the initial word on the tape, i.e., $q_0^0=(s,b)$, $q_1^0=(p^0,x^1)$, $q_i^0=(s,x^i)$ for $2\leq i\leq n$ and $q_{n+1}^0=(s,b)$. For $i\in \comp$ let $\to_i$ be the union of the following transitions.
    \begin{enumerate}
      \renewcommand{\labelenumi}{\alph{enumi})}
      \item \label{onCell}For all $\gamma,\gamma'\in\Gamma$ and $p\in P\setminus\{p^Y,p^N\}$ let $(p,\gamma)\xrightarrow{(p,\gamma)_i^1}_i(s,\gamma')$ if there are $p'\in P$ and $T\in\{-1,1\}$ such that $\delta(p,\gamma)=(p',\gamma',T)$.
      \item \label{toCell}For all $\gamma,\tilde{\gamma}\in\Gamma$, $p\in P\setminus\{p^Y,p^N\}$ and $p'\in P$ let $(s,\tilde{\gamma})\xrightarrow{(p,\gamma)_i^2}_i(p',\tilde{\gamma})$ if there are $\gamma'\in\Gamma$ and $T\in\{-1,1\}$ such that $\delta(p,\gamma)=(p',\gamma',T)$.
    \end{enumerate}

    The transitions described in a) model the impact of $\delta$ on cell $i$ if the tape head is on this cell. Let $M$ be in state $p$ and the tape head on cell $i$ reading $\gamma$, i.e., $T_i$ is in the state $(p,\gamma)$. If $\delta(p,\gamma)=(p',\gamma',T)$ then  $\gamma'$ is written and the tape head moves to a neighboring cell, i.e., $T_i$ moves to the state $(s,\gamma')$. On the other hand, the transitions described in b) model a head movement onto cell $i$.  Let $\tilde{\gamma}$ be the current tape symbol on cell $i$, i.e., $T_i$ is in state $(s,\tilde{\gamma})$ before the head moves. After the movement let $M$ change its state to $p'$, i.e., $T_i$ moves to the state $(p',\tilde{\gamma})$.
    
    \begin{remark}
      $\sys_M$ satisfies the conditions of an interaction system: every port of a component occurs in at least one interaction. Let $i\in \comp$, $(p,\gamma)^1_i\in A_i$ and $\delta(p,\gamma)=(p',\gamma',T)$ then $0\leq i+T\leq n+1$ and $\{(p,\gamma)^1_i,(p,\gamma)^2_{i+T}\}\in \int$. For $(p,\gamma)^2_i\in A_i$ is $0\leq i-T\leq n+1$ and $\{(p,\gamma)^1_{i-T},(p,\gamma)^2_i\}\in \int$.
      
      $\sys_M$ has a linear communication structure because every component $1\leq i\leq n$ only interacts with its neighboring components $i-1$ and $i+1$.
    \end{remark}
    
    \begin{remark}
      The reduction is polynomial in the size of an underlying DTM $M=(\Gamma,\Sigma,P,\delta)$, since $|\int|\leq |P|\cdot|\Gamma|$ and for all $i\in \comp$ $|A_i|\leq 2\cdot|P|\cdot|\Gamma|$ and $|Q_i|\leq (|P|+1)\cdot|\Gamma|$.
    \end{remark}

    \begin{theorem}
      Let $M=(\Gamma,\Sigma,P,\delta)$ be a linear bounded DTM, $x\in\Sigma^*$ with $|x|=n$ an input for $M$ and $\sys_M$ the associated linear interaction system. $M$ accepts $x$ iff a global state $q=(q_0,\dots,q_{n+1})$ is reachable in $\sys$ such that there is $i\in\{0,\dots,n+1\}$ with $q_i=(p^Y,\gamma)$ for a tape symbol $\gamma\in\Gamma$.
      \begin{proof}
        We prove this theorem by giving an isomorphism, with respect to transitions in $\sys_M$ and transitions among configurations in $M$, between global states of $\sys_M$ and configurations of $M$. The statement of the theorem then follows by induction as the isomorphism maps the initial configuration of $M$ to the initial state of $\sys_M$.

        Let $R$ be the set of configurations of $M$. We map $(p;\gamma_0,\dots,\underline{\gamma_i},\dots,\gamma_{n+1})\in R$ to a global state $q=(q_0,\dots,q_{n+1})$ such that $q_i=(p,\gamma_i)$ and $q_j=(s,\gamma_j)$ for $j\not=i$. Let $Q'$ be the set of global states that correspond to the configurations in $R$. It is clear that this mapping is a bijection between $R$ and $Q'$.

        Let $(p;\gamma_0,\dots,\underline{\gamma_i},\dots,\gamma_{n+1})\in R$ and $q=(q_0,\dots,q_{n+1})\in Q'$ be the associated state in $\sys_M$. Let $\delta(p,\gamma_i)=(p',\gamma_i',T)$, i.e., the next configuration in $M$ is $(p';\gamma_0,\dots,\gamma_i',\underline{\gamma_{i+1}},\dots,\gamma_{n+1})\in R$ if $T=1$ (the case $T=-1$ is treated analogously). The only enabled port in component $i$ is $(p,\gamma_i)_i^1$, then the only enabled interaction in $q$ is $\{(p,\gamma_i)_i^1,(p,\gamma_i)_{i+T}^2\}$. Thus, component $i$ reaches the state $(s,\gamma_i')$ and component $i+T$ the state $(p',\gamma_{i+T})$. The resulting global state $q'$ corresponds to the respective configuration in $M$. The fact that the inverse of the mapping is also a homomorphism can be shown analogously.
      \end{proof}
    \end{theorem}

    \begin{remark}
      An instance of the reachability problem is an interaction system $\sys$ and a global state $q$. The interaction system $\sys_M$ for a linear bounded DTM $M$ and an input $x$ can be extended such that a distinguished global state is reached if $M$ halts on $x$. This can be achieved by a technique that is used in \cite{orgtreelike} for tree-like interaction systems. The idea is to invoke, starting from the component that reached $(p^Y,\gamma)$, that each component shall reach a distinguished state. This invocation can be propagated through neighboring components.
    \end{remark}

  \section{PSPACE-completeness of Reachability in Star-Like Systems}\label{star}

    Here we show that deciding the reachability problem in the class of star-like interaction systems is PSPACE-complete by providing a reduction from a general interaction systems $\sys$ to a star-like systems $\sys'$. The idea of the reduction is to construct a ``control component'' $cc$ that forms the center of the star structure in $\sys'$ and is surrounded by the components of $\sys$. An interaction in $\sys$ is modeled by multiple interactions in $\sys'$. The execution of an interaction in $\sys$ then corresponds to the execution of a sequence of interactions in $\sys'$ that is coordinated by $cc$ and achieved in two steps. Let $\alpha$ be an interaction in $\sys$.
    \begin{enumerate}
      \renewcommand{\labelenumi}{\alph{enumi})}
      \item In a first step $cc$ interacts with each component that participates in $\alpha$ and checks whether the respective port in $\alpha$ is enabled without changing the local states of the components. If this check fails then $cc$ returns to its initial state.
      \item If the check succeeds then $cc$ interacts with each respective component on the ports in $\alpha$, i.e., a global transition in $\sys$ that is labeled by $\alpha$ is simulated.
    \end{enumerate}
    Let $Q=\prod_{i\in \comp}Q_i$ be the global state space of $\sys$ then we have a global state space $\prod_{i\in \comp\cup\{cc\}}Q_i$ for $\sys'$ with the property that $q\in Q$ is reachable in $\sys$ iff a state $q'$ is reachable in $\sys'$ such that $q'$ equals $q$ up to the local state of the component $cc$. Since reachability in general interaction systems is PSPACE-complete, the consequence of this transformation is the PSPACE-completeness of reachability in star-like interaction systems.

    Let $\sys=(\im,\{T_i\}_{i\in \comp})$ be an interaction system with interaction model $\im=(\comp,\{A_i\}_{i\in \comp},\int)$ and $\sys'=(\im',\{T_i'\}_{i\in \comp'})$ be an interaction system with interaction model $\im=(\comp',\{A_i'\}_{i\in \comp'},\int')$.

    Let $\comp'=\comp\cup\{cc\}$, where $cc$ is a control component that coordinates sequences of interactions in $\int'$ that correspond to interactions in $\int$. For $i\in K$ let $A_i'=A_i\cup\{a_i^{ok},a_i^{\neg ok}|a_i\in A_i\}$. $a_i^{ok}$ respectively $a_i^{\neg ok}$ models that component $i$ is in a local state that enables respectively does not enable the port $a_i\in A_i$. The set of ports $A_{cc}$ of component $cc$ is given by
    \begin{displaymath}
      A_{cc}=\{a\_i_{cc}^{ok},a\_i_{cc}^{\neg ok},a\_i_{fire}^{cc}|i=1,\dots,n,a_i\in A_i\}\cup \{\alpha_{cc}|\alpha\in \int\}.
    \end{displaymath}
    Let $i\in K$ and $a_i\in A_i$ a port in $i$ then $a\_i_{cc}^{ok}$ models that component $i$ currently enables $a_i$ and $a\_i_{cc}^{\neg ok}$ models that $a_i$ is currently not enabled by $i$. $a\_i_{fire}^{cc}$ models that component $i$ performs a transition labeled by $a_i$. For an interaction $\alpha\in\int$ the port $\alpha_{cc}$ models the initiation of a process that checks whether $\alpha$ is enabled by the respective components and, if applicable, coordinates that all ports in $\alpha$ interact one after another.
    
    The set of interactions $\int'$ is given by
    \begin{eqnarray*}
      \int' &=& \{\{a_i^{ok},a\_i_{cc}^{ok}\},\{a_i^{\neg ok},a\_i_{cc}^{\neg ok}\},\{a_i,a\_i_{cc}^{fire}\}|a_i\in A_i,i=1,\dots,n\}\cup  \{\{\alpha_{cc}\}|\alpha\in \int\}.
    \end{eqnarray*}
    
    The local behavior of $i\in \comp$ is given by $T_i'=(Q_i,A_i',\to_i',q_i^0)$ with
    \begin{displaymath}
      \begin{array}{ll}
        \to_i'= & \to_i\cup \{(q_i,a_i^{ok},q_i)|q_i\in Q_i\wedge a_i\in en(q_i)\}\cup  \{(q_i,a_i^{\neg ok},q_i)|q_i\in Q_i\wedge a_i\notin en(q_i)\}.
      \end{array}
    \end{displaymath}
    
    $T_i'$ extends $T_i$ such that for each port $a_i\in A_i$ there is a loop on each state $q_i\in Q_i$ that is labeled by $a_i^{ok}$ if $q_i$ enables $a_i$ and by $a_i^{\neg ok}$ otherwise. These transitions are used to check whether or not each port of an interaction $\alpha\in\int$ is enabled in a global state of $\sys'$ without changing the local state of the respective components.
    
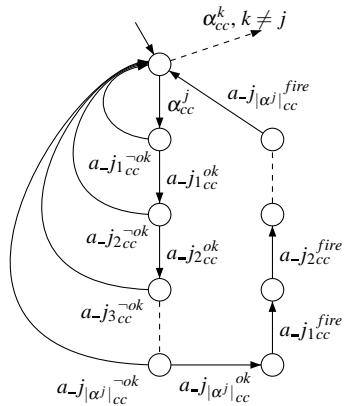
\begin{figure}[H]
      \begin{center}
        \setlength{\unitlength}{1.0cm}
        \begin{picture}(1.0,4.75)
          \gasset{Nw=0.3,Nh=0.3,Nmr=3,AHnb=1}
          \node(A)(1.0,4.0){}
          \imark[iangle=120,ilength=0.5](A)
          \node(B)(1.0,3.0){}
          \node(C)(1.0,2.0){}
          \node(D)(1.0,1.0){}
          \node(E)(1.0,0.0){}
          \node(F)(2.5,0.0){}
          \node(G)(2.5,1.0){}
          \node(H)(2.5,2.0){}
          \node(I)(2.5,3.0){}
          \node[Nframe=n](J)(2.5,4.5){}

          \drawedge(A,B)                      {\scriptsize $\alpha^j_{cc}$}
          \drawedge(B,C)                      {\scriptsize $a\_{j_1}_{cc}^{ok}$}
          \drawedge(C,D)                      {\scriptsize $a\_{j_2}_{cc}^{ok}$}
          \drawedge[dash={0.075}0,AHnb=0](D,E){}
          \drawedge[ELside=r](E,F)            {\scriptsize $a\_{j_{|\alpha^j|}}_{cc}^{ok}$}

          \drawbpedge[ELpos=15](B,180,1.0,A,180,1.0){\scriptsize $a\_{j_1}_{cc}^{\neg ok}$}
          \drawbpedge [ELpos=9](C,180,2.0,A,170,1.0){\scriptsize $a\_{j_2}_{cc}^{\neg ok}$}
          \drawbpedge [ELpos=7](D,180,3.0,A,160,1.0){\scriptsize $a\_{j_3}_{cc}^{\neg ok}$}
          \drawbpedge [ELpos=7](E,180,4.0,A,150,1.0){\scriptsize $a\_{j_{|\alpha^j|}}_{cc}^{\neg ok}$}

          \drawedge[ELside=r](F,G)                  {\scriptsize $a\_{j_1}_{cc}^{fire}$}
          \drawedge[ELside=r](G,H)                  {\scriptsize $a\_{j_2}_{cc}^{fire}$}
          \drawedge[dash={0.075}0,AHnb=0](H,I)      {}
          \drawedge[ELside=r,ELdist=0,ELpos=20](I,A){\scriptsize $a\_{j_{|\alpha^j|}}_{cc}^{fire}$}

          \drawedge[dash={0.075}0,ELpos=80,ELdist=-0.1](A,J)      {\scriptsize $\alpha^k_{cc}$, $k\not=j$}
        \end{picture}
      \end{center}

      \caption{Parts of the behavior of component $cc$.}
      \label{cc}
    \end{figure}

    Let $\alpha^j=\{a_{j_1},\dots,a_{j_{|\alpha^j|}}\}\in\int$. Figure \ref{cc} depicts the part of the local behavior $T_{cc}=(Q_{cc},A_{cc},\to_{cc},q_{cc}^0)$ of component $cc$ that coordinates a test that checks whether each port in $\alpha^j$ is enabled in $\sys'$ and, if applicable, enables ports that can interact with each port in $\alpha^j$. $q_{cc}^0$ is marked by an incoming arrow.
    
    \begin{remark}
      Each port of $\sys'$ occurs in at least one interaction, i.e., $\sys'$ satisfies the conditions of an interaction system. It is clear that $\sys'$ is star-like because each component that originated from $\sys$ interacts only with the control component $cc$.
      
      Furthermore, the size of $\sys'$ is polynomial in the size of $\sys$. $|K'|=|K|+1$, $|\int'|=|\int|+\sum_{i\in K}3\cdot|A_i|$ and for $i\in K$ holds $|A_i'|=3\cdot|A_i|$ and $|\to_i'|=|\to_i|+|Q_i|\cdot|A_i|$. For $cc\in K'$ holds $|A_{cc}|=|\int|+\sum_{i\in K}3\cdot|A_i|$, $|Q_{cc}|=1+\sum_{\alpha\in\int}2\cdot|\alpha|$ and $|\to_{cc}|=\sum_{\alpha\in\int}(3\cdot|\alpha|+1)$.
    \end{remark}

    \begin{theorem}
      Let $\sys$ be an interaction system with components $K$ and $\sys'$ the associated star-like interaction system. A global state $q$ is reachable in $\sys$ iff a global state $q'$ is reachable in $\sys'$ such that $q_i=q_i'$ for $i\in K$ and $q_{cc}'=q_{cc}^0$.
      \begin{proof}
        Let $q$ be a state in the global behavior $T$ of $\sys$ and $q'$ be the state in the global behavior $T'$ of $\sys'$ where $q_i=q_i'$ for $i\in K$ and $q_{cc}'=q_{cc}^0$, i.e., component $cc$ is in its initial state. Consider $\alpha^j=\{a_{j_1},\dots,a_{j_{|\alpha^j|}}\}\in \int$ such that each port in $\alpha^j$ is enabled in $q$, i.e., all local states $q_l'$, $l=j_1,\dots,j_{|\alpha^j|}$ in $q'$ enable the ports $a_l^{ok}$ and $a_l$ and do not enable $a_l^{\neg ok}$. $q'$ enables the interaction $\{\alpha_{cc}^j\}$. If this interaction is performed then the only possible sequence of interactions results in a state $\tilde{q}'$ with $\tilde{q}_i=\tilde{q}_i'$ for $i\in K$ and $\tilde{q}_{cc}'=q_{cc}^0$. Let there be a port in $\alpha^j$ that is not enabled in $q$, e.g., $q_l$ with $l\in\{j_1,\dots,j_{|\alpha^j|}\}$ does not enable $a_l$ then $q_l'$ does enable $a_l^{\neg ok}$ and not $a_l^{ok}$. If $\{\alpha_{cc}^j\}$ performed in $q'$ then the only possible sequence of interactions in $\sys'$ leads back to state $q'$. For the global initial states $q^0$ of $\sys$ and ${q^0}'$ of $\sys'$ holds that $q^0_i={q^0_i}'$ for $i\in K$ and ${q^0_{cc}}'$ is the initial state of the local behavior of component $cc$. The ``if'' part follows by induction over paths in the global behavior of $\sys$. The ``and only if'' part follows analogously.
      \end{proof}
    \end{theorem}

  \section{Conclusion and Related Work}\label{con}
  
      We investigated complexity issues for classes of interaction systems that are relevant in various applications. One with a linear the other with a star-like communication pattern. We showed that even for these simply structured systems deciding the reachability problem is PSPACE-complete. These results strengthen PSPACE-completeness results of the reachability problem in general interaction systems \cite{everythingIsPSPACE}. The formalism of interaction systems is very basic, and thus our results are easily applicable to other formalisms that model cooperating systems. Our results justify techniques that are based on a sufficient condition and establish reachability or reachability dependent system properties in subclasses of cooperating systems that are defined by a restricted communication structure that forms a star or a line or in respective superclasses, which are sketched in the following.
      
      \cite{bernardoArch} examined a process algebra based on an architectural description language called \textit{PADL} and considers deadlock-freedom in systems with a tree-like communication pattern (a proper superclass of systems with a star-like or linear pattern). The technique is based on a compatibility condition that is tested among pairs of cooperating subsystems, i.e., the composite behavior of two subsystems is weak bisimilar to the behavior of one of the components. An efficient technique based on a sufficient conditions for establishing deadlock-freedom in interaction systems with a star-like communication pattern is introduced in \cite{Lambertz09} where, similar to \cite{bernardoArch}, a compatibility condition based on branching bisimilarity is tested. A sufficient condition for establishing deadlock-freedom for the subclass of tree-like interaction systems is described in \cite{treeLike} where a condition is tested on the reachable state spaces of pairs of interacting components. In \cite{Lambertz11} the condition in \cite{treeLike} is extended such that deadlock-freedom can be established in a proper superclass of tree-like interaction systems. Hennicker et al. proposed in \cite{BHH+06,HJK08} a technique to construct so called \textit{observable} behavior of a cooperating system with an acyclic communication pattern which can be used to establish certain system properties. \cite{roscoeArch} describes a general communication graph for CSP models and shows how tree structures can be constructed by merging several processes. \textit{Communicating Sequential Processes} are introduced in \cite{hoare} where a directed communication structure based on input/output communication is considered. It is argued that communicating processes, if a directed input/output communication structure forms a rooted tree, can not deadlock.
      
    \nocite{*}
    \bibliographystyle{eptcs}
    \bibliography{biblio}
    
\end{document}